\documentclass[aps,pre,floatfix,twocolumn,footinbib,showpacs]{revtex4-1} 
\usepackage{epsfig}
\usepackage{hyperref}
\usepackage{color,soul}
\usepackage{multirow}
\usepackage{ulem}
\usepackage{cancel}
\definecolor{orange}{rgb}{1,0.5,0}
\definecolor{dorange}{rgb}{0.5,0.25,0}
\definecolor{violet}{rgb}{1,0,1}
\definecolor{lsmblue}{rgb}{0.4,0.7,1}
\definecolor{dlsmblue}{rgb}{0.2,0.35,0.5}
\definecolor{pink}{rgb}{1,0.7,0.7}
\definecolor{teal}{rgb}{0.6,1,1}
\definecolor{dgreen}{rgb}{0,0.5,0}
\begin{document}
\title{Dusty plasma cavities: probe-induced and natural}
\author{B. J. Harris}
\email[Electronic mail: ]{brandon\_harris@baylor.edu}
\author{L. S. Matthews}
\email{lorin\_matthews@baylor.edu}
\author{T. W. Hyde}
\email{truell\_hyde@baylor.edu}
\affiliation{CASPER (Center for Astrophysics, Space Physics, and Engineering Research)\\Baylor University, Waco, Texas 76798-7310, USA}
\date{\today}
\begin{abstract}

A comprehensive exploration of regional dust evacuation in complex plasma crystals is presented. Voids created in 3D crystals on the International Space Station have provided a rich foundation for experiments, but cavities in dust crystals formed in ground-based experiments have not received as much attention. Inside a modified GEC RF cell, a powered vertical probe was used to clear the central area of a dust crystal, producing a cavity with high cylindrical symmetry. Cavities generated by three mechanisms are examined. First, repulsion of micrometer-sized particles by a negatively charged probe is investigated. A model of this effect developed for a DC plasma is modified and applied to explain new experimental data in RF plasma. Second, the formation of natural cavities is surveyed; a radial ion drag proposed to occur due to a curved sheath is considered in conjunction with thermophoresis and a flattened confinement potential above the center of the electrode. Finally, cavity formation unexpectedly occurs upon increasing the probe potential above the plasma floating potential. The cavities produced by these methods appear similar, but each are shown to be facilitated by fundamentally different processes.

\end{abstract}


\maketitle
\section{\label{sec:Intro} Introduction}

Voids in complex plasma are well known phenomena occurring in microgravity experiments conducted aboard parabolic flights \cite{butten} and the International Space Station (ISS) \cite{kretsch}. However, the related structures often seen in ground-based experiments have been the subject of more limited study. 

The 3D voids observed in the PKE-Nefedov experiments have been explained employing an ion drag force directed spherically outward from the center of the bulk plasma, balanced by an inward electric field force. Analogous to the 3D void observed on the ISS, a central region devoid of dust is often observed in ground-based experiments where the dust particles, levitated in the sheath electric field, form planar crystalline structures. In this paper, such circular 2D regions devoid of dust will be referred to as ``cavities." Ion streaming alone cannot explain the natural formation of these dust cavities, since streaming ions flow essentially vertically through the sheath toward the lower electrode. Such naturally formed cavities have been explored under specific plasma conditions \cite{paeva}, and a numerical simulation using the ion momentum equation attributed void formation to an increased central outward radial electric force \cite{hu}.

In order to better understand the natural formation of cavities, in this paper a vertically aligned biased probe was used to create a dust-free region within a dust crystal, as shown in Fig.~\ref{void}. This technique is similar to experiments performed in a DC plasma in which dust was repelled from a horizontally aligned probe \cite{thomas}. The electric field from a negatively biased probe repels the dust particles, increasing the cavity radius as the bias is further decreased. Surprisingly, increasing the probe potential above the floating potential also increases the cavity size, as shown in Fig.~\ref{radcurr}. 

This paper analyzes cavities created using two different methods, a negative/positive probe bias, and cavities formed naturally when plasma conditions are altered to reduce the overall horizontal confinement. It will be shown that different mechanisms are responsible for cavity formation in each of these cases.

In section \ref{sec:Neg}, a numerical model similar to that employed by Thomas et al. \cite{thomas} to analyze cavities is used to find the electric field of the negative probe, showing that the local electron density at the dust particle height must exhibit a small increase with decreasing probe potential. 

In section \ref{sec:Nat}, it is shown by an analysis of the potential energy that natural cavities can be produced by a flattening of the central horizontal confinement potential and thermophoresis, enhanced slightly by curvature of the lower sheath edge.

Finally in section \ref{sec:Pos}, for a positively charged probe, results from previous work \cite{harris} are utilized to show that a radial outward force (resulting from an increase in gravitational potential energy as the sheath edge is raised) coupled with an outward ion drag (due to ion repulsion by the probe) can also generate a cavity.

\section{\label{sec:Exp} Experiment}

\subsection{\label{sec:App} Apparatus}

This experiment employed a modified GEC (Gaseous Electronics Conference) RF reference cell \cite{hargis, land} at the Center for Astrophysics, Space Physics, and Engineering Research (CASPER). The cell contains two parallel plate electrodes, 8 cm in diameter and displaced from one another by 1.9 cm; the lower is powered at a frequency of 13.56 MHz while the upper is ring-shaped and grounded. In the experiment, 8.9 micron diameter spherical melamine formaldehyde (MF) dust are levitated in an argon plasma. Plates with milled 1 inch diameter cylindrical depressions of 1 and 3 mm were placed atop the lower electrode to provide horizontal confinement of the dust. Two 60 fps cameras were employed to obtain top and side views of the particles, which were illuminated by Coherent Helium Neon lasers.

\sethlcolor{teal}
\begin{figure}
\includegraphics[scale=0.63,trim=2 0 0 0]{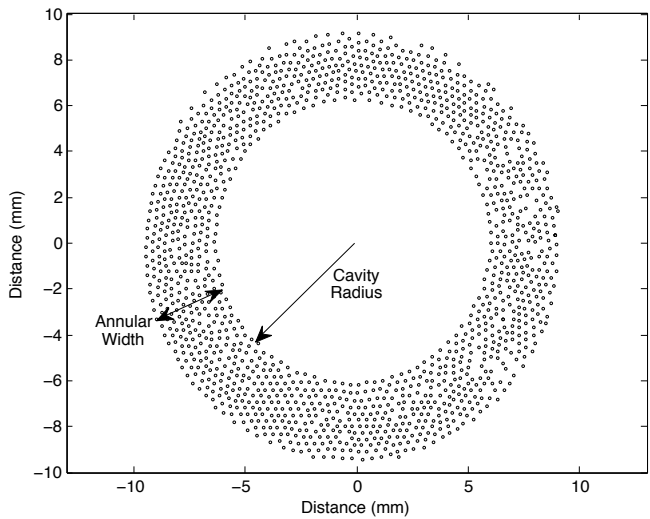}
\caption{Dust particle positions in a planar dust crystal with an open central cavity created by a negatively biased vertical probe located at the center of the cavity. The system power was 10 W for a neutral gas pressure of 100 mTorr, providing a probe potential of $-30$ V. A 1-inch diameter circular cutout with depth of 1 mm was used for horizontal confinement.
\label{void}} 
\end{figure}

A Zyvex S100 head mounted within the cell provided physical manipulation of a powered probe. The S100 head allows remote controlled movement of up to 10 mm in all three dimensions with nanometer precision.  A 48 mm long cylindrical probe was attached to the head, which tapered from 500 microns in diameter to 250 microns over 100 $\mu$m. A power supply was connected to the probe allowing application of user-defined potentials to be applied. Additional information regarding the probe can be found in Ref.~\cite{harris}.

\sethlcolor{orange}
\begin{figure}
\includegraphics[scale=0.61,trim=2 0 0 0]{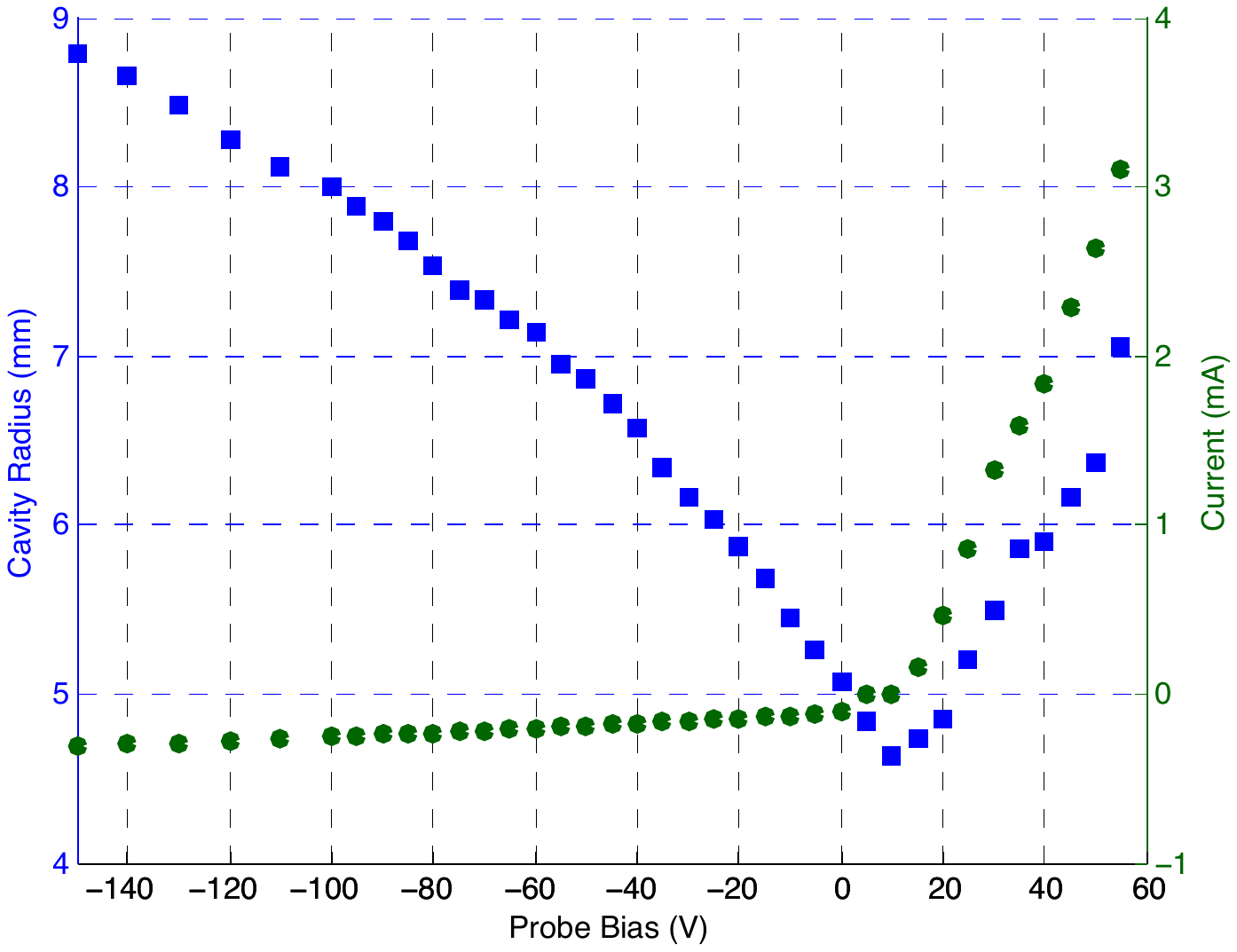}
\caption{(Color online) The probe potential-induced cavity size (blue squares) as a function of probe bias with respect to ground. System parameters are the same as those in Fig. \ref{void}. The probe current (green circles) is superimposed, showing the significant difference between negative and positive probe biases.
\label{radcurr}} 
\end{figure}

\subsection{\label{sec:Neg} Negative Potential Probe-induced Cavities}

In order to directly create dust cavities, the vertical probe was lowered through the upper electrode and the plasma bulk. Data was collected at pressures of 50, 100, and 300 mTorr, and system powers of 1, 5 and 10 W, while the probe bias was varied over the range 0 to $-50$ V. The tip of the probe was positioned to intersect the plane of the dust crystal. This was achieved in all cases except those at system powers of 5 and 10 W with background pressure of 300 mTorr; for these cases, the dust levitation height was found to change significantly as a function of the neutral gas pressure. Since cavity size was not found to be exceedingly sensitive over the actual range of tip position relative to the dust crystal (for example, at 10 W and 100 mTorr, the cavity radius increased by 0.6 mm when the probe tip was lowered 2 mm below the crystal plane, and decreased by 0.9 mm when raised 3 mm above), this will not be considered an issue.

A detailed survey of cavity size as a function of probe bias for RF power of 10 W and gas pressure of 100 mTorr, is shown in Fig.~\ref{radcurr}. The floating potential is defined by the point where the current to the probe goes to zero, here at 8.9 V; the current to the probe for potentials higher than this increases much faster than it decreases for potentials below the floating potential. The rapid increase of current to the probe above the floating potential may be attributed to the fact that the more mobile electrons respond more quickly to changes in potential than do the heavier ions. The current varies linearly with the probe bias in both regions. Note that higher positive probe potentials ($>$ 55 V) stimulated arcing; therefore, an electron saturation asymptote could not be reached due to the large surface area of the probe, preventing it from functioning as a Langmuir probe.

\begin{figure}
\includegraphics[scale=0.583,trim=2 0 0 0]{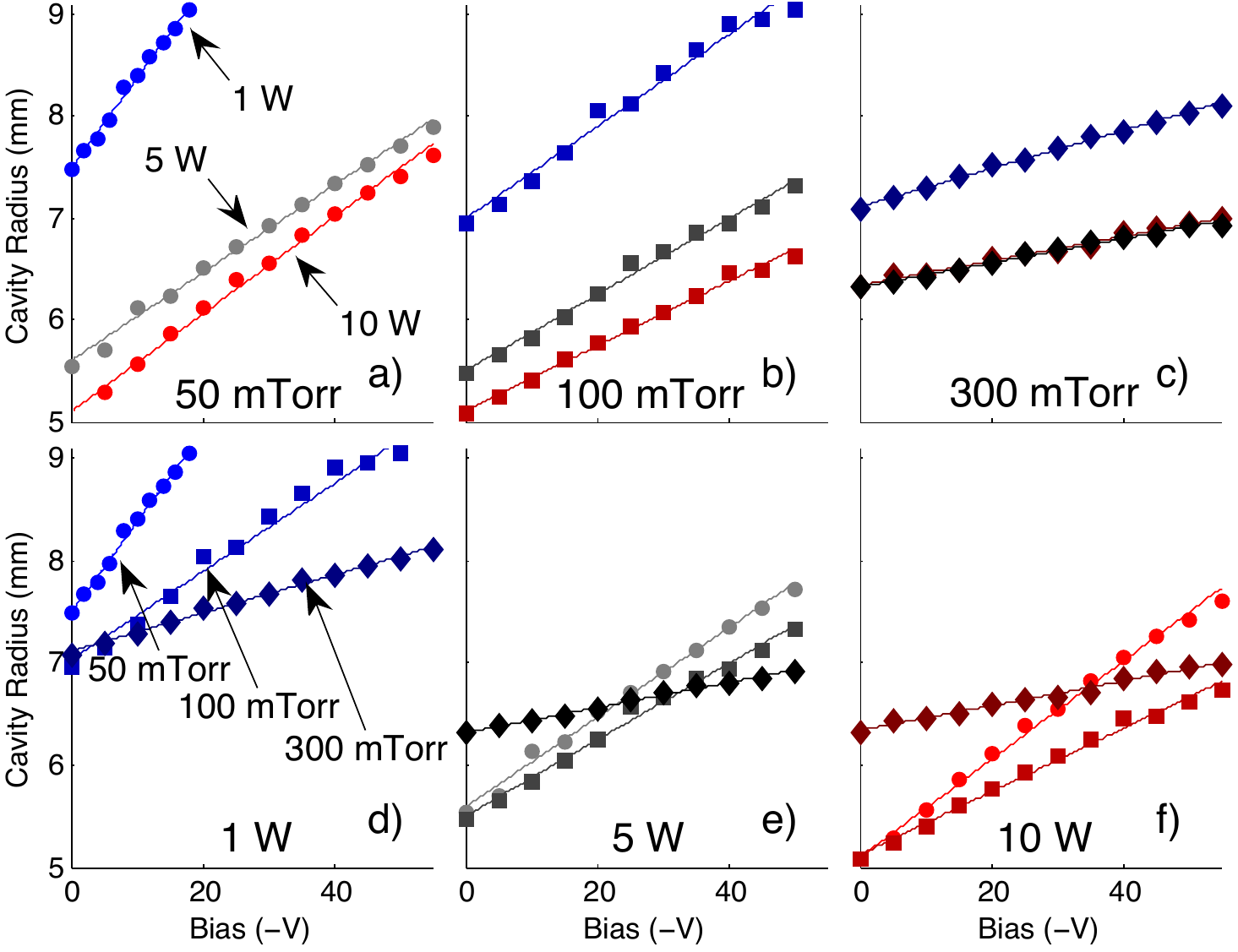}
\caption{(Color online) Cavity radii versus probe bias. In plots a-c) system pressure is fixed as indicated, for powers of 1 W (blue), 5 W (gray) and 10 W (red).  In d-f) results are shown for a constant power as indicated, with the circles denoting a pressure of 50 mTorr, the squares 100 mTorr, and diamonds 300 mTorr. Lines represent linear fits to the data.\label{negresults}}
\end{figure}

A series of plots in Fig.~\ref{negresults} shows experimentally produced cavity sizes, increasing linearly with decreasing (more negative) probe potential. At lower pressures (\ref{negresults}a), ion collisions are limited so sensitivity to the probe potential is increased (as reflected in larger slopes). At higher pressures (\ref{negresults}c), the range of cavity size is reduced (as reflected in decreased slopes), even though the cavities at 0 V bias are larger (shown later to be the result of a natural cavity contribution). At lower power (\ref{negresults}d) cavity size at 0 V bias is also larger since the overall ionization is smaller, which makes the screening length longer. The reverse is true for higher power, though again the natural cavity takes effect at P = 300 mTorr (\ref{negresults}f).

\sethlcolor{orange}

A model previously developed by Thomas et al.~\cite{thomas} for numerically calculating the dust cavity size in a DC plasma was adapted to the 10 W, 100 mTorr case shown here. A balance between ion drag directed inward and the probe electric force directed outward is assumed. In this experiment more dust is present, so an interparticle repulsion (assuming Yukawa interactions) was included as an inward radial force on particles at the cavity edge ($F_{ion}+F_E+F_{Y,r}$ = 0). The form of the ion drag used here is given in Ref.~\cite{khrapak2},
\begin{equation}
\label{khrapakiondrag}
\begin{array}{c} F_i=\sqrt{2 \pi} r_p^2 n_i m_i v_{Ti}^2 \left\{ \sqrt{\frac{\pi}{2}} erf\left( \frac{u}{\sqrt{2}} \right) (1+u^2+ \right. \\ \left. (1-u^{-2})(1+2z\tau) + 4z^2\tau^2u^{-2}ln(\Lambda))+ \right. \\ \left. u^{-1}(1+2z\tau+u^2-4z^2\tau^2ln(\Lambda))exp\left( -\frac{u^2}{2} \right) \right\}, \end{array}
\end{equation}
where $r_p$ is the dust grain radius, $n_i$ is the ion density, $m_i$ is the ion mass, $v_{Ti}$ is the ion thermal speed, $u$ is the ion streaming speed normalized to $v_{Ti}$, $z$ is the normalized grain charge, $\tau$ is the ratio of electron to ion temperature, and $\Lambda$ is the Coulomb logarithm, defined as $(z\tau r_p/\lambda_D u^2+1)/(z\tau r_p/\lambda_D u^2+r_p/\lambda_D)$, where $\lambda_D$ is the screening length, defined as $(1/\lambda_{De}^2+1/\lambda_{Di}^2(1+u^2))^{-1/2}$. The electric field as a function of radius is found (by taking the negative derivative of the potential) from a solution of the Poisson equation,
\begin{equation}
\label{thomaspoisson}
\phi''(r) + \frac{en_{i0}}{\epsilon_0}\left( \frac{E_0}{|\phi'(r)|}-\frac{n_{e0}}{n_{i0}}e^{e\phi(r)/kT_e} \right)=0,
\end{equation}
where $\phi$ is the electric potential, $kT_e$ is the electron energy, and $E_0$ is the electric field at the cavity edge. Finally, the full Yukawa force is
\begin{equation}
\label{yukawa}
F_Y=\frac{Q^2}{4 \pi \epsilon_0 x}\left( \frac{1}{x}+\frac{1}{\lambda_D} \right) e^{-x/\lambda_D},
\end{equation}
where $Q$ is the grain charge, and $x$ is the distance between two dust grains. The radial component of $F_Y$ is extracted to generate $F_{Y,r}$. 

\sethlcolor{orange}
\begin{figure}
\includegraphics[scale=0.585,trim=2 0 0 0]{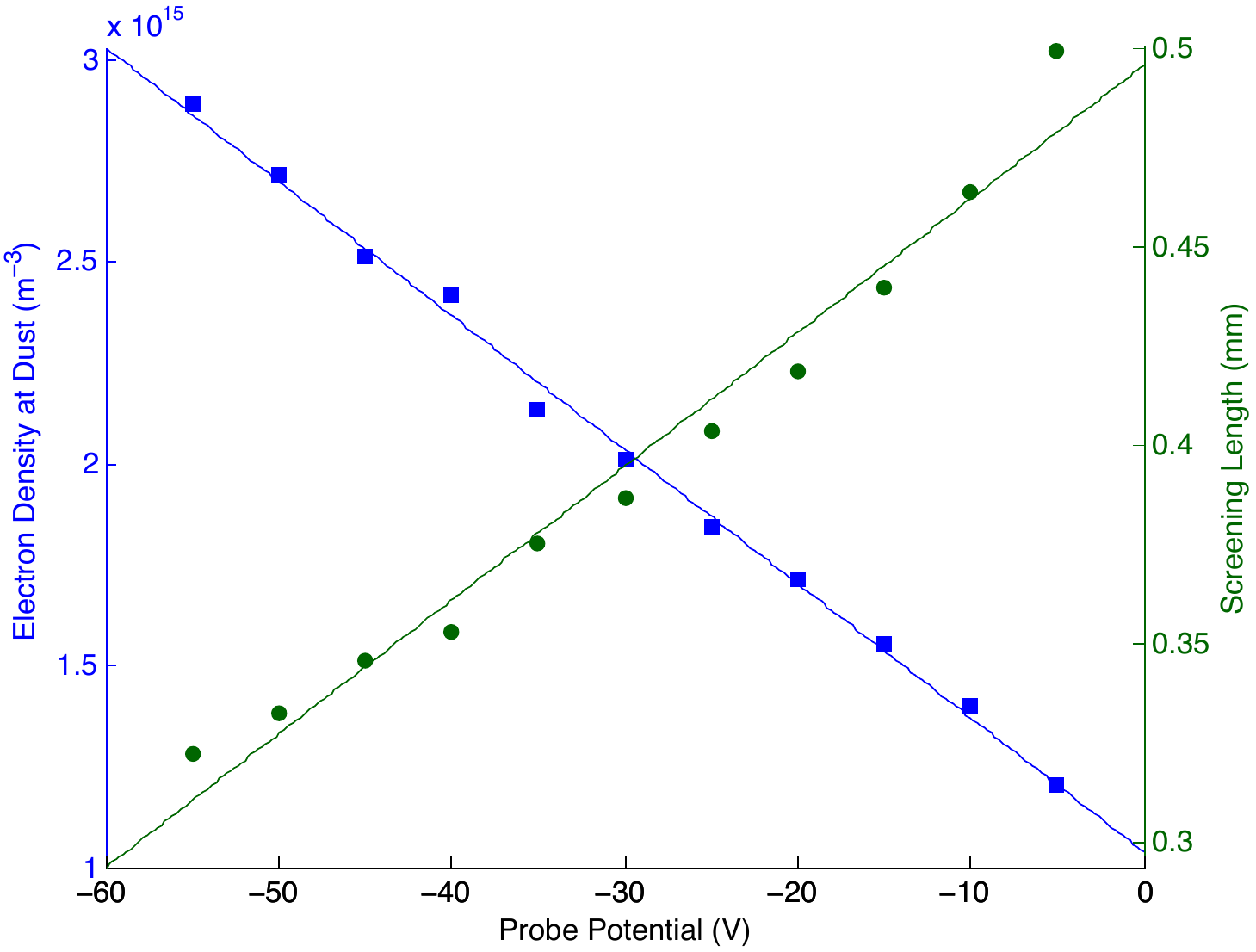}
\caption{(Color online) Electron density at the dust as a function of probe potential. The data shown was determined by fixing the cavity radius to the experimental value and adjusting the density until the model-predicted potential at the probe matched the experimentally imposed value. As shown, the change in electron density causes a change in the plasma screening length (green circles). The lines represent linear fits.
\label{nechange}}
\end{figure}

In order for the model to match the measured cavity radius, the electron density at the dust levitation height must be adjusted linearly with probe bias, as shown in Fig.~\ref{nechange}. This increase in electron density at the cavity edge under a decrease in probe potential may be explained by considering that electrons repelled from the probe region are still confined within the plasma. By extrapolating the fit to the floating potential (8.9 V), the electron density at the dust can be determined, providing an alternate measurement method to a Langmuir probe. Note that in Eqn.~\ref{thomaspoisson}, electric potential is relative to the plasma potential, which was 34 V with respect to ground for this case. The charge (calculated using OML theory) and the screening length also depend on the electron density. Linear fits were used to find their equilibrium values of 49,000e and 530 $\mu m$, respectively.
\subsection{\label{sec:Nat} Natural Cavities}
\subsubsection{\label{sec:Obs} Observations}

\begin{figure}
\includegraphics[scale=0.63,trim=2 0 0 0]{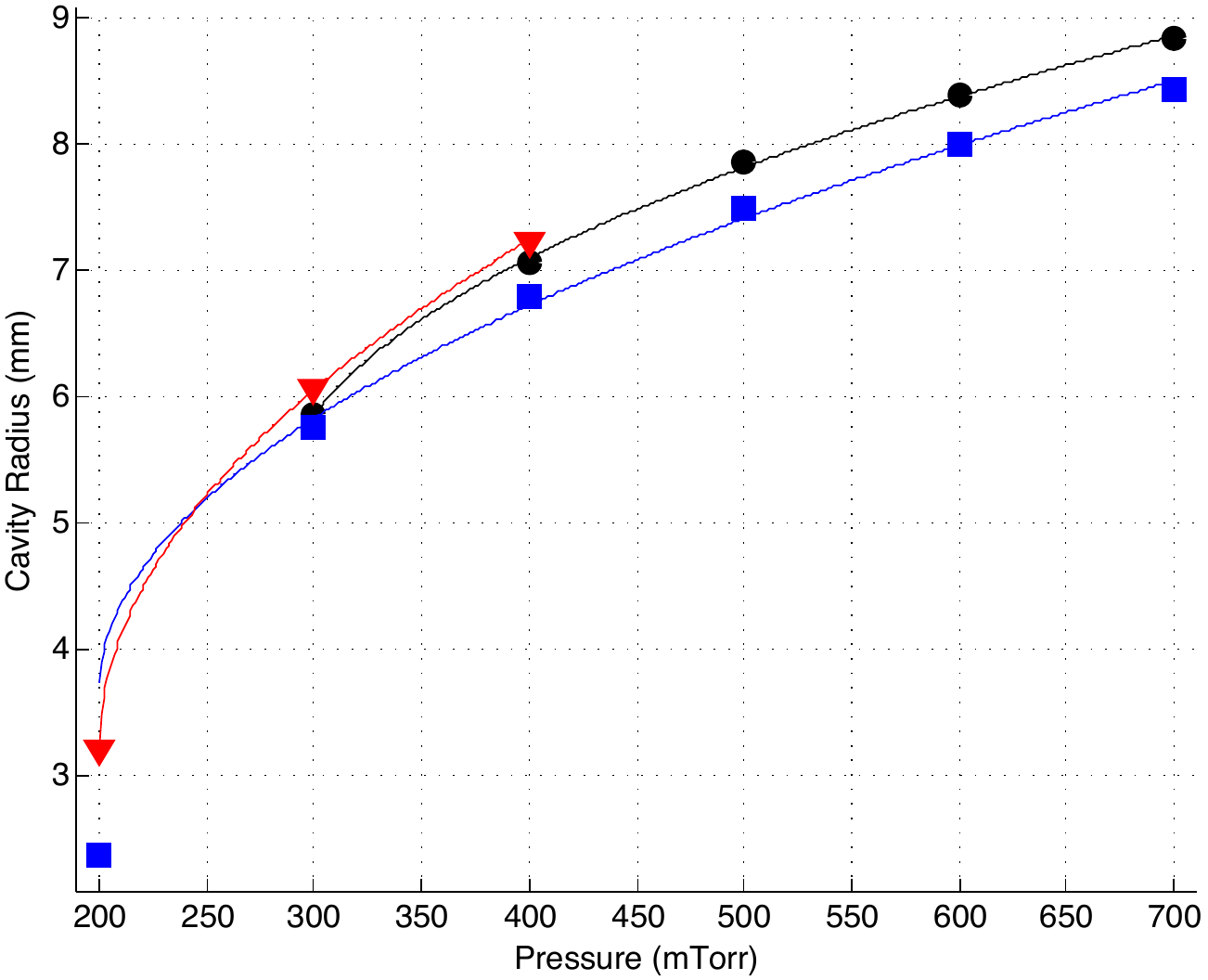}
\caption{(Color online) Natural cavity radius versus pressure for 10 W system power. The (blue) squares and (black) circles indicate cavity radii found using a 1 mm lower electrode cutout at both natural bias and a fixed DC bias of $-50$ V, respectively. Using the fixed bias did not permit plasma ignition at 200 mTorr. The (red) triangles represent cavities produced using a 3 mm lower electrode cutout for horizontal confinement, as in Ref.~\cite{paeva}. At pressures greater than 400 mTorr, the dust levitated within the 3 mm cutout, preventing it from being illuminated by the horizontal laser.
\label{prescav}}
\end{figure}

Natural cavities were found to appear in the crystal at high system powers $(\ge$ 10 W) and to grow in size when the system pressure was increased ($>$ 200 mTorr). To explore this behavior, the system pressure was varied at a set power of 10 W to produce a cavity, as shown in Fig.~\ref{prescav}. This effect was first described by Paeva et al.~\cite{paeva}, who used a 3 mm deep cutout to provide horizontal confinement. In the current set of experiments, cavities formed within crystals confined by both 1 and 3 mm deep cutouts were essentially the same size. In either case, lower pressures resulted in smaller cavity size, consistent with previous results \cite{paeva2}. The fit lines to the cavity size as shown in Fig.~\ref{prescav} are of the form $r$ $\propto$ $P^{1/2}$, where $P$ is the pressure difference with respect to the pressure at the cavity's appearance (200 mTorr). Note that a similar functional dependence on pressure is seen in voids created under microgravity conditions in dust crystals \cite{lipaev}, where significant outward ion drag and thermophoresis was also found to generate natural voids \cite{land2}.

\sethlcolor{yellow}

\begin{figure}
\includegraphics[scale=0.63,trim=2 0 0 0]{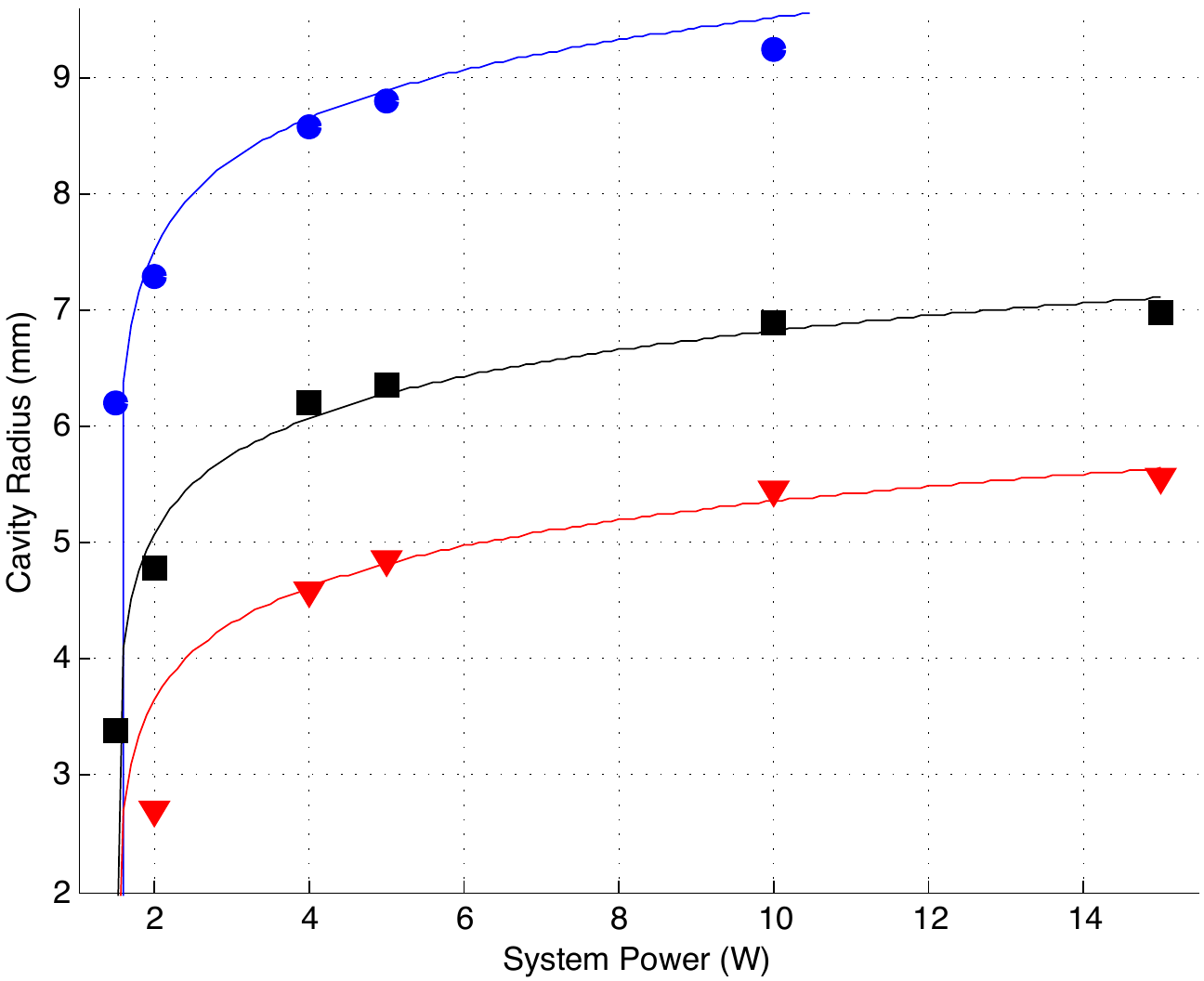}
\caption{(Color online) Natural cavity radius versus system power for various fixed pressures.  The (blue) circles represent cavities found using the 1 mm deep lower electrode cutout depression at 750 mTorr, whereas the (black) squares and (red) triangles represent cavities found using the 3 mm deep electrode cutout, which was also used by Ref.~\cite{paeva}, at 350 mTorr and 250 mTorr, respectively.
\label{powcav}}
\end{figure}

Natural cavities were also found to develop at higher pressures ($\ge$ 250 mTorr) when the power is increased, as shown in Fig.~\ref{powcav}. Initial cavity growth is much more sensitive to changes in power, with best fits found for $\gamma$ $<$ 1/2 in the relation $r$ $\propto$ $P_o^\gamma$.

\begin{figure}
\includegraphics[scale=0.593,trim=2 0 0 0]{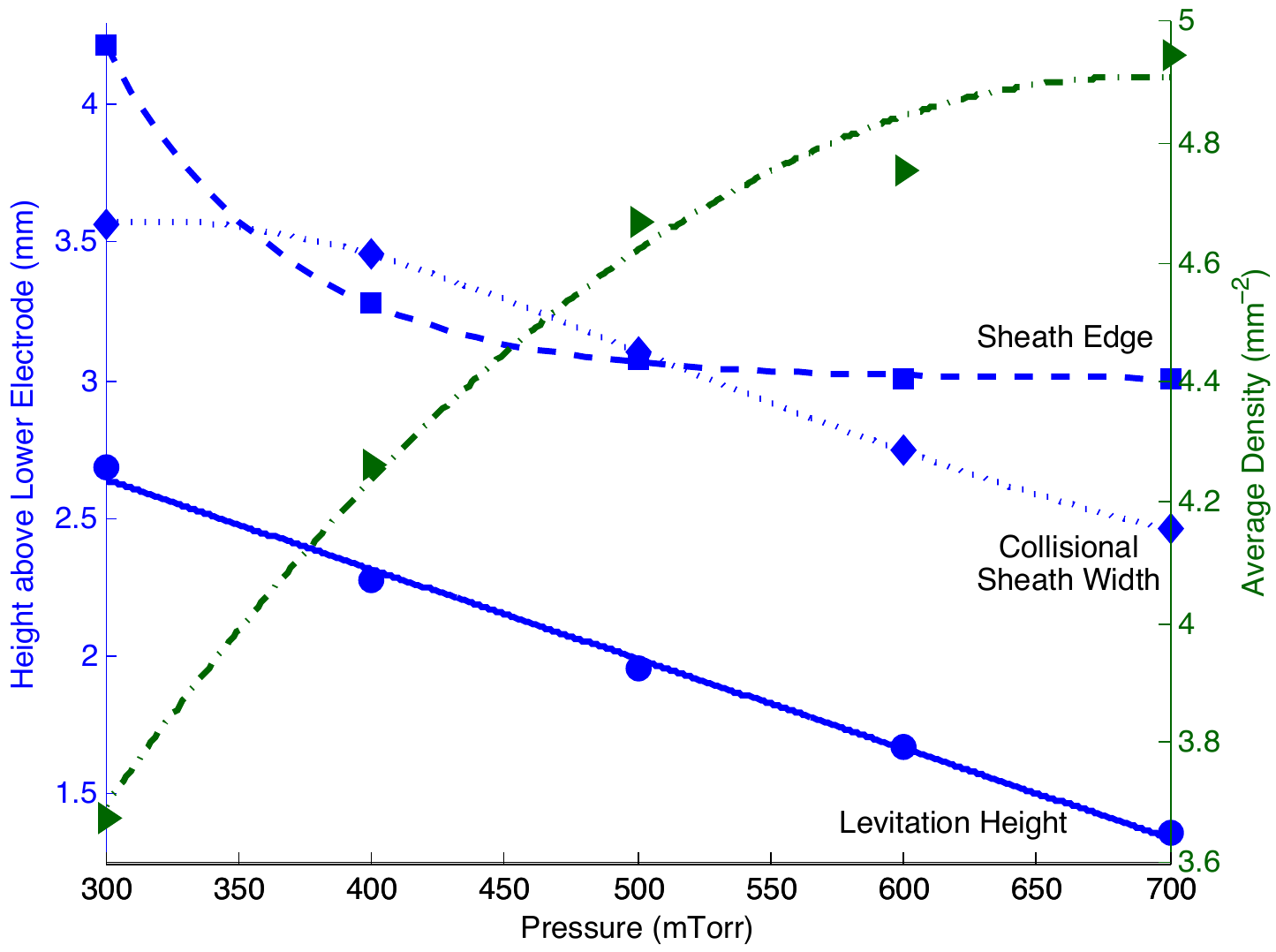}
\caption{(Color online) Distance from the lower electrode as a function of pressure. The experimentally determined sheath edge (squares with dashed fit) decreases nonlinearly with increasing pressure, whereas the dust levitation height (circles with solid fit) decreases linearly. A theoretical collisional sheath width (diamonds with dotted fit) also decreases with pressure, as well as coinciding with the experimental sheath width (edge), showing the applicability of the collisional regime. The average areal density of the dust (green triangles with dash-dot fit), which was found by taking the particle number and dividing by the area of the dust annulus, increases as the pressure is increased.
\label{sidemeasures}}
\end{figure}

As the pressure increases, the sheath edge and the levitation height of the particles decrease (Fig.~\ref{sidemeasures}). The levitation height decreases linearly, while the sheath edge (defined by the levitation height of 0.46 micrometer nanoparticles as in Ref.~\cite{samarian}) asymptotically approaches a minimum value. A collisional sheath width derived in Ref.~\cite{sheridan} agrees with the experimentally determined sheath edge.

\sethlcolor{orange}

The ratio of the equilibrium dust height to the sheath edge position (shown in Table \ref{parameters}) was found to decrease from 69 to 45 percent as the pressure increased from 400 to 700 mTorr. The slight increase from 300 to 400 mTorr can be explained by the small increase in plasma densities over the same pressures, despite the levitation height decreasing. The mean free path and screening lengths for each pressure are also reported in Table \ref{parameters}. The ion and electron densities at the dust position are found by modifying the values found in the bulk, while requiring continuity and energy conservation of the ions, and a Boltzmann distribution for the electrons,
\begin{equation}
n_{i0}u_b=n_i u_i
\label{continuity}
\end{equation}
\begin{equation}
\frac{1}{2}m_i u_i^2=\frac{1}{2}m_i u_b^2-eV
\label{energycons}
\end{equation}
\begin{equation}
n_e=n_{e0}e^{eV/kT_e}.
\end{equation}
Here $n_{i0,e0}$ is the initial bulk ion (electron) density, $u_b$ is the assumed ion Bohm speed at the sheath edge, $u_{i}$ is the ion speed, $V$ is the electric potential with respect to the plasma potential, and $T_e$ is the electron temperature, as employed in Ref.~\cite{harris}.

\subsubsection{\label{sec:Prev} Previous Explanations for Cavities}

\sethlcolor{orange}

Several theories have previously been proposed to explain the formation of natural cavities \cite{goree, hu, paeva, paeva2}. Samsonov and Goree proposed that at high powers ($\geq$ 100 W), natural cavities are produced due to an instability caused by a local change in ionization created by electron depletion from the plasma to the dust cloud (i.e., Havnes parameter of 1-2) \cite{samsonov}. The total number of electrons in our plasma equals $n_eV=n_e A_e \Delta S=$ 2.1$\times 10^{14}$, where $A_e$ is the area of the electrode and $\Delta S$ is the plasma width, which is much greater than the total number of electrons on the dust, $4.6\times 10^6$ (calculated using the electron density and particle number for a pressure of 300 mTorr shown in Table \ref{parameters}). Given this and the fact that we have an open system, electron depletion to the dust may be considered to be negligible.

\sethlcolor{orange}
\begin{table}
\begin{tabular}{|l|ccccc|}
\hline\hline
\multicolumn{1}{|c}{Parameter} & \multicolumn{5}{|c|}{Pressure (mTorr)}\\
\hline
\multicolumn{1}{|c|}{System Power = 10 W} & 300 & 400 & 500 & 600 & 700\\
\hline
Particle Number & 451 & 423 & 393 & 323 & 311\\
Sheath Edge ($mm$) & 4.2 & 3.3 & 3.1 & 3.0 & 3.0\\
Dust Height / Sheath Edge (\%) & 64 & 69 & 64 & 56 & 45\\
Sheath Shift ($\mu m$) & 280 & 310 & 330 & 360 & 390\\
Confinement Energy ($10^{-12}$ J) & 3.1 & 3.0 & 2.8 & 2.4 & 2.3\\
Ion Density ($10^{15}$ $m^{-3}$) & 1.3 & 1.4 & 1.2 & 1.0 & 0.8\\
Electron Density ($10^{15}$ $m^{-3}$) & 0.7 & 1.0 & 0.7 & 0.4 & 0.1\\
Electron Temperature (eV) & 8.7 & 8.8 & 9.0 & 9.3 & 9.0\\
Mean Free Path ($\mu m$) & 160 & 120 & 90 & 73 & 61\\
Screening Length ($\mu m$) & 660 & 580 & 700 & 960 & 1500\\
Dust Charge ($10^3e$) & 10 & 8.8 & 8.1 & 8.2 & 8.6\\
Repulsive Energy ($10^{-14}$ J) & 4.9 & 3.2 & 3.5 & 3.2 & 4.3\\
Radial Drag (\%) & 0.24 & 0.24 & 0.25 & 0.28 & 0.31\\
Zero Point ($mm$) & 6.3 & 7.0 & 7.5 & 7.9 & 8.0\\
\hline\hline
\end{tabular}
\caption{Experimental parameters found for natural cavity conditions. The radial ion drag is listed as a percentage of the gravitational force on a dust grain of radius 4.45 $\mu m$. The zero point is defined as the radial distance at which the horizontal confinement potential energy was found to be zero before increasing quadratically in order for the location of the minimum potential energy to equal the average particle distance from the electrode center (see Fig.~\ref{pote}). Energies are listed as totals for all particles. The screening length and plasma densities are reported at the dust levitation height.}
\label{parameters}
\end{table}

Cavity formation has also been attributed to other mechanisms. Recently, Hu et al. identified an increased central radial electric field through numerical simulation, and a reduced central radial confinement as one possible cause, although they did not consider ion drag \cite{hu}. Alternatively, Paeva et al. included radial ion drag due to the curvature of the sheath edge, but did not include the reduced confinement \cite{paeva}. Later, numerical modeling (also by Paeva et al.) including a flat-bottomed cutout, used for horizontal confinement in all experiments, confirmed a non-zero radius of curvature for the sheath edge \cite{paeva2}.

\subsubsection{\label{sec:New} New Explanation}

In the present work, natural cavity formation is explained by analyzing the potential energy of the dust particles. This offers an advantage over the model in Section \ref{sec:Neg}, which cannot be used without a probe to find the electric field in the cavity. The potential energies involved include the interparticle repulsive ($PE_r$), the ion ($PE_{ion}$), the thermophoretic ($PE_{therm}$), and the confinement ($PE_{conf}$). $PE_r$ is given by
\begin{equation}
\sum_{dust}{\frac{Q^2}{8\pi\epsilon_0 r_d}e^{-r_d/\lambda_D}},
\end{equation}
where $r_d$ is the separation between each pair of particles. $PE_{ion}$ is determined from the radial component of the ion drag force (Eqn.~\ref{khrapakiondrag}) of the streaming ions. $PE_{therm}$ is estimated from the force of the radial temperature gradient. $PE_{conf}$ is calculated from the gravitational potential energy difference between levitation heights for particles inside and outside the electrode depression (see Table \ref{parameters}). Since $PE_r$, $PE_{ion}$, and $PE_{therm}$ direct particles parallel to the lower electrode while gravity acts perpendicularly, the only radial escape for a particle is to move upward, out of the cutout, making energy the more useful quantity versus force. Coupling these with flattened confinement, the potential energies are sufficient to produce the observed natural cavity sizes, which will be shown by finding the minimum of their sum.

As the pressure is increased, the interparticle potential energy decreases initially due to a decrease in charge, but increases again primarily because of an increase in areal density of the dust (Fig.~\ref{sidemeasures}). The dust particle number decreases with increased pressure, signifying that $PE_{conf}$ is exceeded by the sum of other potential energies. However, the horizontal confinement potential has typically, as a first approximation, been assumed to be parabolic \cite{zhang}. At higher pressures, as the particles approach the lower electrode, the amount of plasma between the electrode and dust is too thin to provide a smooth horizontal electric field transition all the way to the electrode center, due to the discontinuous electrode surface (from high outside the cutout to low inside). 

\begin{figure}
\includegraphics[scale=1,trim=0 0 0 0]{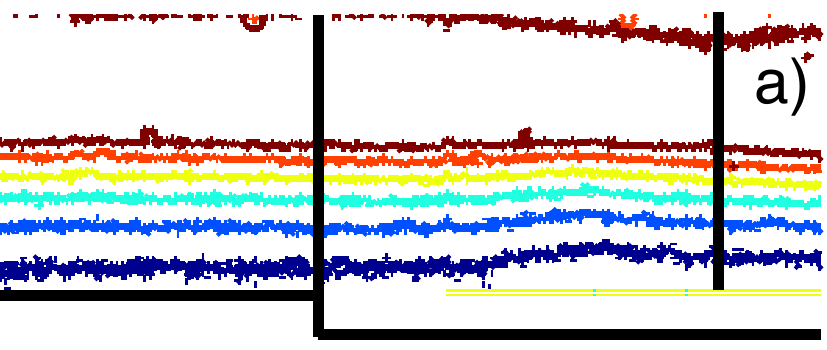}\\
\includegraphics[scale=0.97,trim=0 10 2 3]{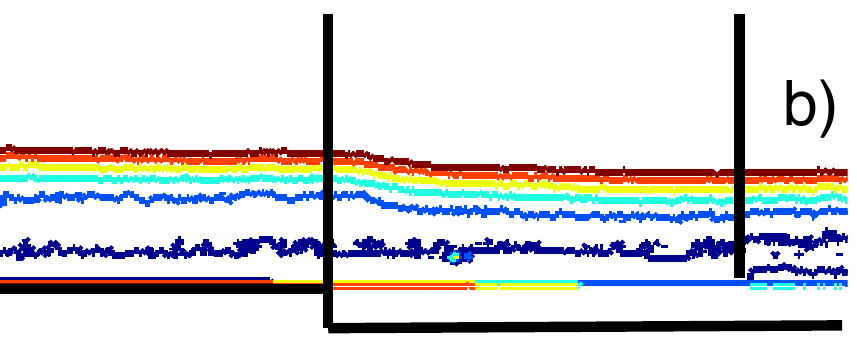}
\caption{(Color online) Side-view plasma intensity contours above the lower electrode at a) 200 mTorr and b) 700 mTorr. The contour line nearest the bulk (burgundy) shows visible curvature in the high pressure case. The heavy black line indicates the shape of the electrode surface, with the vertical lines indicating the cutout edge (left) and center (right).\label{contours}}
\end{figure}

The curvature of the sheath edge can be determined by examining the intensity of the plasma glow. As shown in Fig.~\ref{contours}, the curvature becomes more pronounced at higher pressures as the sheath width decreases and the plasma and the particles move closer to the lower electrode. This provides an added level of complexity; not only does the dust get closer to the lower electrode with increased pressure, but the sheath curvature increases as well. It is typically assumed that ions flow in a direction perpendicular to the sheath edge; the average radial component of the ion flow was found by $(\Delta S_e - \Delta S_0)/R$, where $\Delta S_e$ is the sheath width at the cutout edge, $\Delta S_0$ is the sheath width at the cutout center, and $R$ is the radius of the cutout. In order to analyze the radial ion drag, vertical image profiles of small regions (300 microns wide and the height of the image) at the center of the electrode cutout and at the cutout edge were used to determine the sheath edge, defined for this analysis to be the height at which the plasma intensity decreases by a factor of 1/e from the maximum intensity \cite{beckers}. The difference in sheath width ranged from 0 to 400 $\mu m$ as pressure increased from 200 to 700 mTorr (see Table \ref{parameters}) as $\Delta S \propto P^{1/2}$, the same functional relationship as for the cavity size. The levitation of nanoparticles could not be used for determination of the sheath edge at these pressures because the plasma emission intensity increased to such a degree that the nanoparticles could no longer be distinguished.

The most significant contribution to natural cavity formation comes from thermophoresis. Land, Matthews, Hyde, and Bolser recently provided a numerical simulation, which modeled the plasma particles as a fluid and included dust grain dynamics for the conditions used in this experiment. They found an outward radial force on the dust \cite{land2}.

\sethlcolor{orange}

To model $PE_{conf}$ flattening, the gravitational potential energy was set to zero up to a particular radial distance, referred to as the ``zero point." The confinement potential was then increased quadratically, since this provides the simplest (linear) restoring force that increases toward the cutout edge. The zero point was adjusted until the minimum of the total potential energy matched the experimental average annular radius of the dust in the ring, which monotonically increased in size with increased pressure. Decreasing the plasma power decreases the cavity size, so the zero point will drop to 0 mm at lower powers, reducing horizontal confinement to a parabolic potential well, a model that was assumed in Ref.~\cite{zhang} for 1 W. The end point of the horizontal confinement potential energy was set to the gravitational potential energy found from the height of the cutout in addition to the shift of the sheath at the cutout edge ($F_g$ $\times$ (D + $\Delta S_e - \Delta S_0$) = 7.5$\times 10^{-15}$ J for 700 mTorr, where D is the depth of the electrode depression). By considering the average annular radius (effectively the radial center of mass), the repulsive potential energy between dust grains could be ignored. Additionally, the radial component of the interparticle repulsion becomes less significant as the pressure increases due to a decrease in the annular width.

The horizontal confinement depends on the vertical dust position (the shape of the potential well is a function of height), so the dust lattice density is of interest. The areal dust density increases with pressure, as the cavity increases in size, which compresses the dust against the horizontal confinement. The dust density also reaches a maximum limit, as dust is pushed out of the confinement after the total potential energies of the outward forces exceed the inward forces. 

\begin{figure}
\includegraphics[scale=0.643,trim=6 0 0 0]{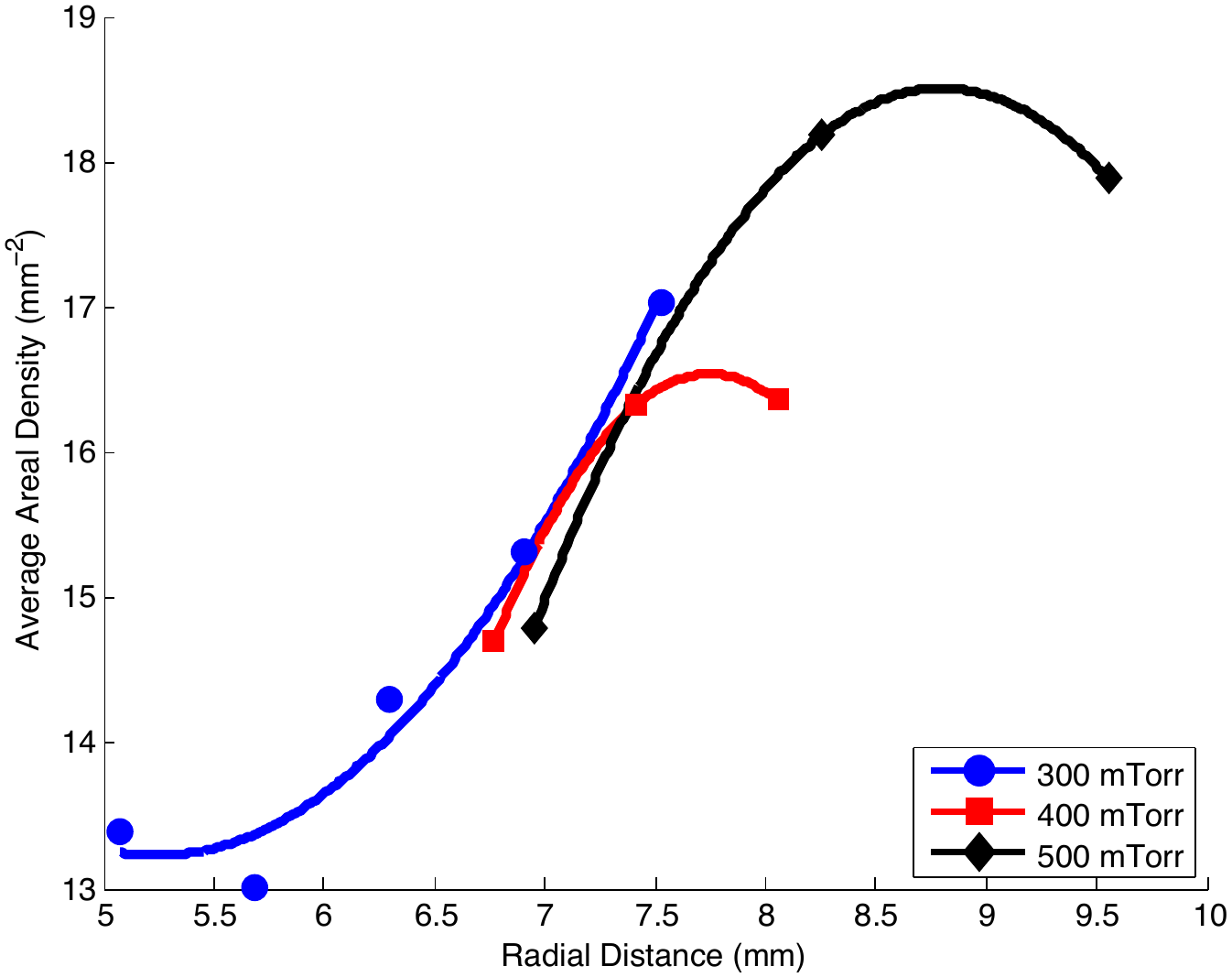}
\caption{(Color online) The average areal dust density as a function of the radial position, where radial distance of dust particles is binned and the interparticle distance averaged. The dust density increases as a function of the radius until it reaches a maximum near the outer edge of the annulus.
\label{raddens}}
\end{figure}

The average dust density as a function of the radial position is also a point of interest in both gravity and microgravity. On the ISS, the highest dust density occurs near the void edge (Fig.~2 in Ref.~\cite{lipaev}), because the ion drag at the void edge increases toward the center of the void faster than the electric field decreases. However, for a cavity, the maximum density is found at the outer edge of the dust annulus, as shown in Fig.~\ref{raddens}. This effect occurs because the total potential energy grows faster at the outer edge of the cavity (due to the electrode cutout edge) than toward the cutout center, as shown below.

\sethlcolor{lsmblue}

The thermophoretic force was calculated through estimation of the change in temperature between the cutout center and edge from the model results and scaling the plasma dimensions in the thermophoresis simulation \cite{land2} to the plasma volume size in this experiment. It was found to be nonzero above a pressure of 100 mTorr, though the model is not applicable below this pressure. Therefore, a thermophoretic force is not significant for the conditions discussed in sections \ref{sec:Neg} and \ref{sec:Pos}, where the pressure was 100 mTorr. For intermediate pressures the force was approximated from a linear fit, and its corresponding potential energy determined by taking the integral of its dot product with the unit tangent vector of the confinement.

\begin{figure}
\includegraphics[scale=0.639,trim=1 0 0 0]{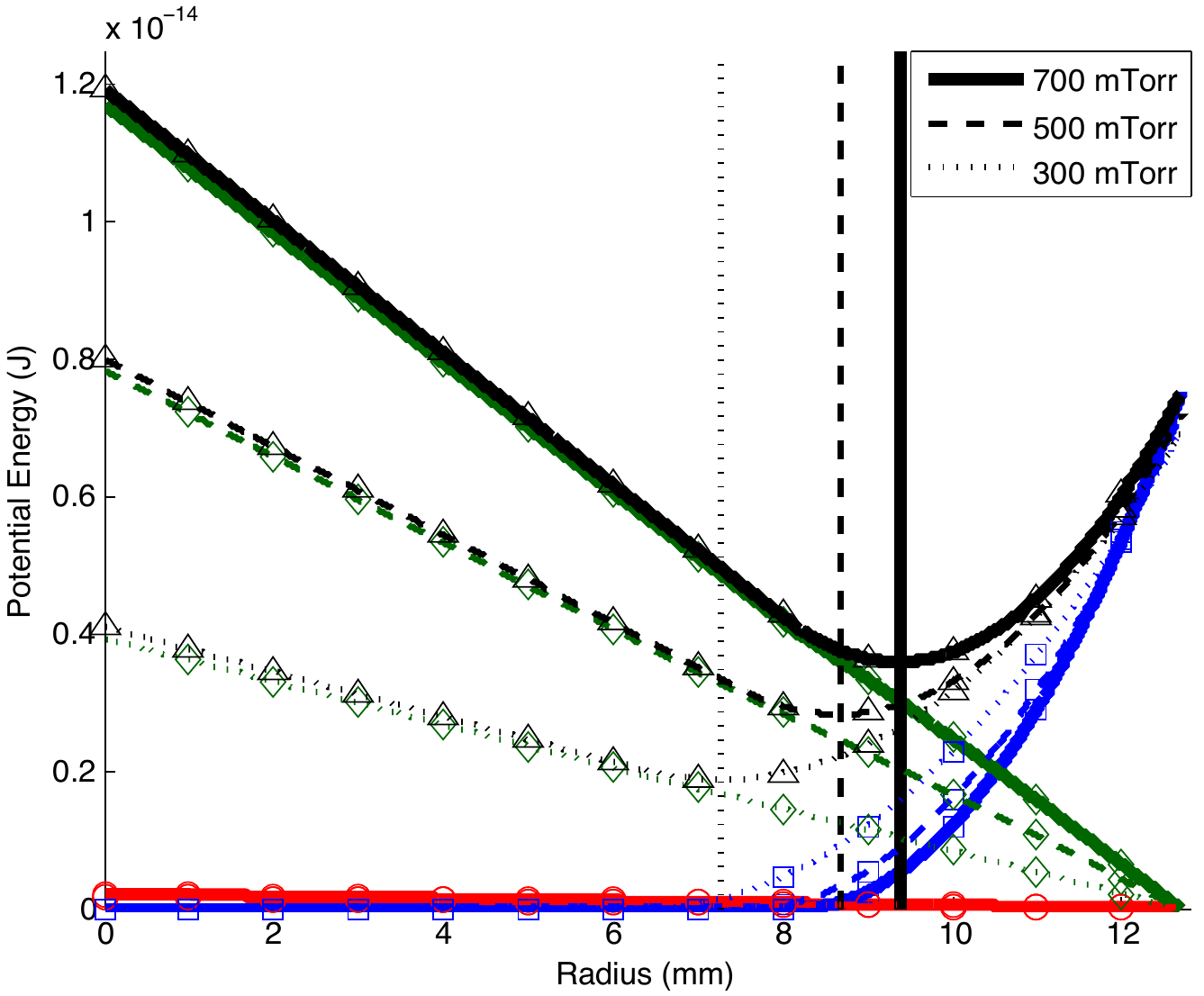}
\caption{(Color online) Potential energy of the average particle in a dust crystal having a naturally formed cavity, shown for various pressures. The (black) vertical lines show the average annular radii, which coincide with the minima of the total potential energy (black triangles). Contributions to the total potential energy include the potential energy of thermophoresis (green diamonds), radial ion drag (red circles), and horizontal confinement (blue squares).
\label{pote}}
\end{figure}

Since the mean free path of the ions decreases with pressure, a collisional model was employed to calculate the dust charge, and in turn the ion drag, at each pressure shown in Table \ref{parameters} \cite{khrapak}. The resultant average potential energy per particle is shown in Fig.~\ref{pote}. Although the radial ion drag force increases with an increase in pressure (see Table \ref{parameters}), the increase of the thermophoretic force and flattening of the confinement provide contributions of much greater magnitude.

In summary, natural cavities form in a single horizontal layer in the plasma sheath under gravity, for increased pressure and/or power. They are important to consider as background contributions to probe-induced cavities when those are generated at conditions favorable for natural cavity emergence.

\subsection{\label{sec:Pos} Positive Potential Probe-induced Cavities}

\sethlcolor{orange}

As shown in Fig.~\ref{radcurr}, cavity size also grows when the probe potential is increased above the floating potential. One difference from negative probe potential-induced cavities is the time required to establish the cavity; positive probe-induced cavities form in seconds as opposed to almost immediately. This is due to the fact that the positive probe takes time to absorb electrons (limited by the plasma-probe surface area) and establish an outward directed ion flow, whereas the negative probe rapidly evacuates the less massive electrons from the probe region, diminishing shielding and allowing the repulsive electric field to affect the dust. This is corroborated by the finding that current to the probe is much greater than that for negative potentials; this implies such cavity growth occurs through an independent mechanism from those described in section \ref{sec:Neg}. Experimental data also shows that the sheath edge directly beneath the probe was raised upon application of positive probe potentials, which may provide an explanation \cite{harris}. The radial expanse of that perturbation was not considered previously, but due to plasma shielding the radial extension of sheath perturbation is expected to be limited, and is modeled in the manner below.

Because ions in plasma are not typically considered to have a Boltzmann distribution around a charged particle \cite{gurnett} and electrons are not mobility limited (the drift velocity is negligible \cite{mcdaniel}), the reverse of the negative probe electric field model (Eqns. 1-3 in \cite{thomas}) could not be applied. Therefore, as in the analysis of natural cavities, the cavity growth was modeled using a potential energy analysis ($PE=PE_{sheath}+PE_{conf}+PE_{ion}$). As a first approximation, it was assumed the potential energy from the raised sheath edge, $PE_{sheath}(r)$, decreases with distance as
\begin{equation}
E_0 e^{-0.5(r/\sigma)^2},
\label{gauss}
\end{equation}
where $r$ is the radial distance, $\sigma$ is a parameter that quantifies the extent of the shielding, and $E_0$ is the gravitational potential energy of a dust particle above the electrode center. The shielding parameter, $\sigma$, was chosen such that the location of the minimum total potential energy matched the average experimental radial dust position ($\sigma$ was found to equal 4.7 $mm$ for all cases), and $E_0$ $=$ $mgh$, where $h$ (the dust levitation height) was found from the amplitudes of the oscillation of the dust at corresponding positive probe potentials as in Ref.~\cite{harris}.

\sethlcolor{teal}
\setstcolor{dgreen}
\begin{figure}
\includegraphics[scale=0.637, trim=2 0 0 0]{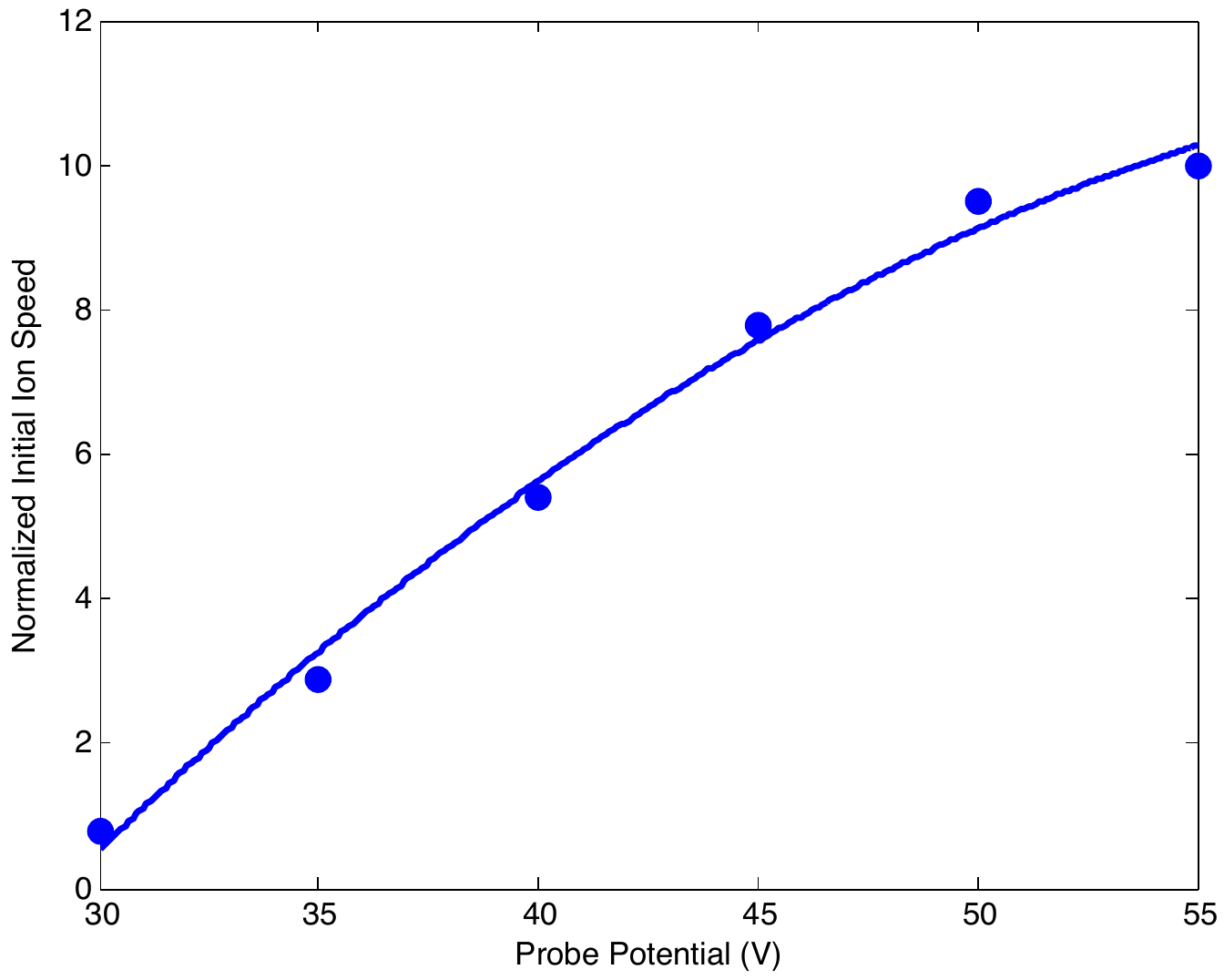}
\caption{Initial ion speed as a function of positive probe potential normalized to the equilibrium ion thermal speed at the dust levitation height. The line is a quadratic fit. The values are found by fitting the minimum of the total potential energy to match the average radial dust position.
\label{iontherm}}
\end{figure}
\setstcolor{dorange}

Since system pressure and power were static, the horizontal confinement potential energy, $PE_{conf}$, was fixed. Its characteristic zero point was found by extrapolating the quadratic fit of the natural cavity zero points (listed in Table \ref{parameters}) to 100 mTorr, resulting in a value of 4.3 $mm$.

\sethlcolor{lsmblue}
\begin{figure}
\includegraphics[scale=0.66,trim=2 0 0 0]{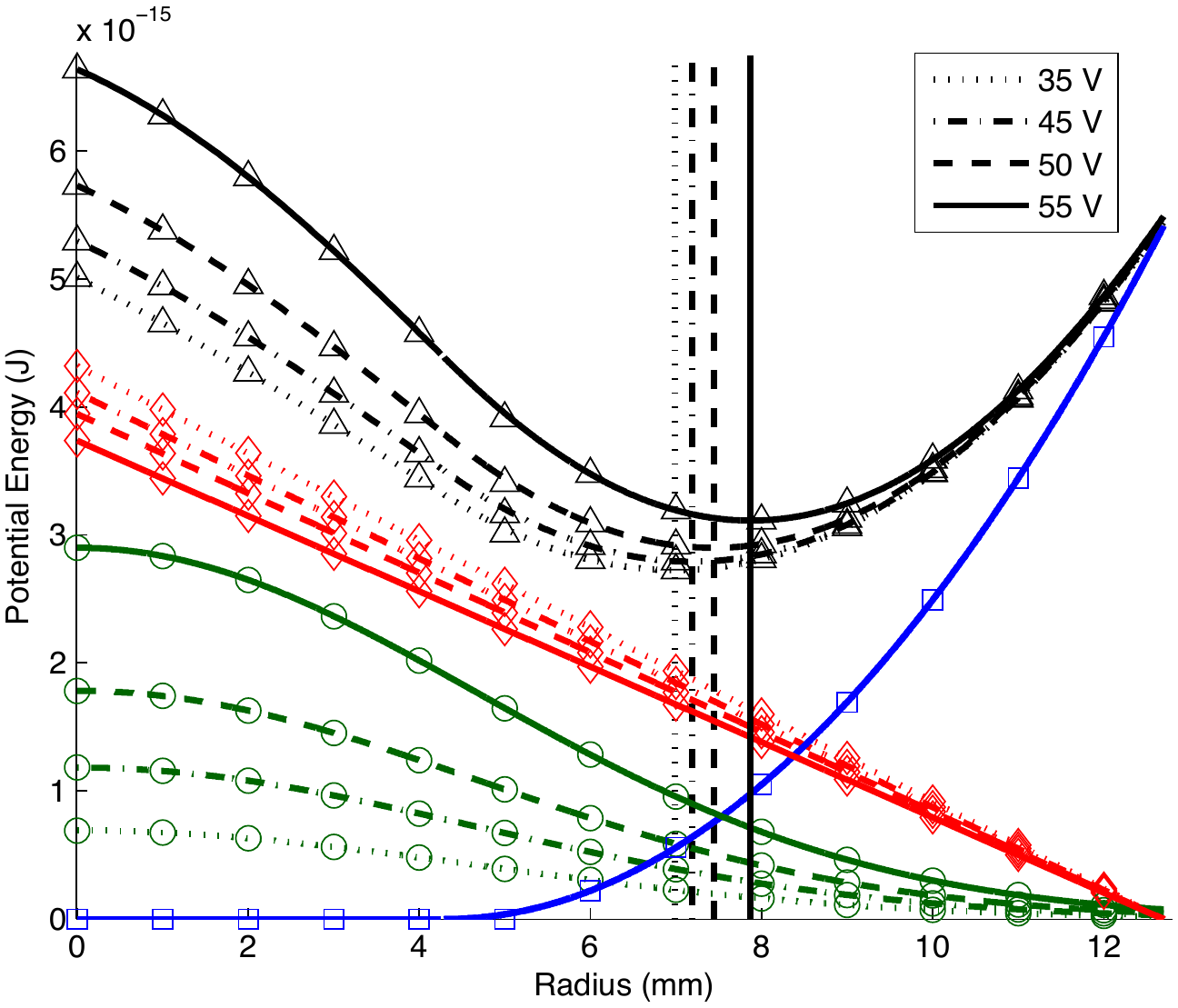}
\caption{(Color online) Potential energy plot for positive-probe-potential induced cavities. The (black) triangles indicate the total potential energy, with contributions from the repulsive energy due to the raised sheath edge (green circles), outward ion drag (red diamonds) and the horizontal confinement (blue squares).  The vertical lines indicate the minimum in the total potential energy, which coincides with the average experimental radial position of the dust. 
\label{potepos}}
\end{figure}

In this case, the potential energy due to the ion flow away from the probe, $PE_{ion}$, plays a significant role. Since the cavities are relatively large compared to the shielding length, it is reasonable to assume that ions reach the sheath plasma potential at the equilibrium levitation height of the dust before impacting the dust. Therefore, energy conservation (Eqn.~\ref{energycons}) may be used to find the velocity of the ions, and the continuity equation (Eqn.~\ref{continuity}) applied to calculate the ion density and solve for the ion drag upon the dust. 

The only remaining unknown parameter is the initial ion velocity at the probe, which will differ from the thermal velocity. The initial ion velocity was left as a free parameter and adjusted to fit the data, as shown in Fig.~\ref{iontherm}. Although it increases from the thermal velocity by a factor of 10 at a probe potential of 55 V, this is still less than the Bohm velocity, which is larger than the thermal velocity by a factor of 13.5. At these potentials, an enhanced glow around the probe suggests new sheath formation, which can justify the ion acceleration. 

The results of the potential energy calculations are shown in Fig.~\ref{potepos}. The ion drag force is found to decrease with an increase in probe potential, but remains larger than the force from the raised sheath. However, the contribution from the raised sheath grows nonlinearly and becomes more important at higher probe potentials.

\section{\label{sec:Con} Conclusions}

Cavity formation in complex plasma crystals has been shown to occur in several forms: induced by a probe charged negatively relative to the plasma, naturally, and induced by a positively charged probe. Each of these was shown to be produced due to a different mechanism, as listed in Table \ref{causes}.

Negative probe potential induced cavities were explained by an electric force directed outward from the probe, balanced with an inward directed ion drag, as also shown in a previous model for probe-induced cavities in a DC plasma \cite{thomas}. In the RF case, interparticle forces had to be included since the annular width of the dust ring (see Fig.~\ref{void}) was greater than the diameter of a single grain. The electron density at the dust was found to vary over the potential range (Fig.~\ref{nechange}), becoming greater for more negative probe potentials. This provides a method to calculate the equilibrium parameters (electron density, screening length, and charge) in the sheath at the dust height by linearly extrapolating the results to the floating potential. Current to the probe varied linearly with probe bias (Fig.~\ref{radcurr}), and consequently the variation of cavity size was linear (Fig.~\ref{negresults}). 

Natural cavities were found to be produced by an outward thermophoretic force and a small radial ion drag, coupled with flattening of the horizontal confinement potential at high pressure (Fig.~\ref{pote}). The radial ion drag was identified by employing plasma emission analysis which showed a small shift in the sheath height between the center and edge of the electrode cutout used for horizontal confinement. The depth of the depression of the electrode was not found to alter the cavity size (Fig.~\ref{prescav}). This may be because the sheath curvature increases just enough to counteract the increased horizontal confinement of the deeper well. The natural cavities examined here differed from voids produced in ISS experiments in that the maximum radial dust density occurred near the cavity edge (Fig.~\ref{raddens}). Low power laboratory cavities also exist in the sheath and not the bulk, making them more than simply 2D void analogs.

\sethlcolor{orange}
\begin{table}
\begin{tabular}{|c|c|}
\hline\hline
\multicolumn{1}{|c|}{Cavity Type} & Creation Mechanisms\\
\hline
Negative Probe & Probe Electric Force\\
\hline
\multirow{2}{*}{Natural} & Thermophoretic Force\\
& Streaming Ion Drag Force\\
\hline
\multirow{2}{*}{Positive Probe} & Probe Generated Ion Drag Force\\
& Central Raised Sheath Edge Barrier\\
\hline\hline
\end{tabular}
\caption{Summary of cavities created and their formation factors in decreasing order of importance.}
\label{causes}
\end{table}

Positive probe potential-induced cavities are perhaps the most interesting because their existence was not expected. Although true that a positive probe attracts negatively charged dust, other effects override that attraction. From a previous experiment (Ref.~\cite{harris}), the sheath edge was found to be raised by a positively charged probe. This forms a central cylindrical potential energy barrier (Fig.~\ref{potepos}) which repels the dust. Simultaneously, the positive probe repels plasma ions, generating an outward radial ion drag force. While ion drag played little role for the negative probe potential-induced cavities, it constitutes the largest generating force in this case.

Future work will employ additional electrode depression depths and radial sizes to further examine whether the natural cavity size remains unaffected; examine the average particle potential energy with a decrease in power to test for a reduction to a parabolic confinement potential; and utilize a polydisperse dust distribution to observe how the dust size affects cavity radius as well as whether this provides a method for radial dust separation by size.

\end{document}